# BAYESIAN BASED COMMENT SPAM DEFENDING TOOL[*]

Dhinaharan Nagamalai[1], Beatrice Cynthia Dhinakaran[2] and Jae Kwang Lee [2]

[1] Wireilla Net Solutions, Australia.

[2] Department of Computer Engineering,Hannam University, Daejeon, South Korea.

## ABSTRACT

*Spam messes up user's inbox, consumes network resources and spread worms and viruses. Spam is flooding of unsolicited, unwanted e mail. Spam in blogs is called blog spam or comment spam.It is done by posting comments or flooding spams to the services such as blogs, forums,news,email archives and guestbooks. Blog spams generally appears on guestbooks or comment pages where spammers fill a comment box with spam words. In addition to wasting user's time with unwanted comments, spam also consumes a lot of bandwidth. In this paper, we propose a software tool to prevent such blog spams by using Bayesian Algorithm based technique. It is derived from Bayes' Theorem. It gives an output which has a probability that any comment is spam, given that it has certain words in it. With using our past entries and a comment entry , this value is obtained and compared with a threshold value to find if it exceeds the threshold value or not. By using this cocept, we developed a software tool to block comment spam. The experimental results show that the Bayesian based tool is working well. This paper has the major findings and their significance of blog spam filter.*

## 1. INTRODUCTION

The growth of the Internet from a mere Local area network to worldwide network plays an important role in day to day activities of humanity. Especially in the field of e-business, e-science, e-publications, the third generation life online is predicted to feature intuitive Artificial Intelligence applications that work shiftily across broadband Internet connections. A web 3.0 target is to develop AI agents that mine mountains of information on the Internet to provide materials that suit the interests of people they serve. The Internet search has become the first choice to gather information leads to a competition to achieve the top ranks from popular search engines. This trend is wide spread in Internet from adult industry to scientific research community. The web world has become a base for billions of dollars worth of business.

---

[*]Earlier version of this paper was published in ISA 2009 Conference,Seoul,Korea [8]

Apart from the information available in the web pages the blogs, forums and email archives have become a source for search engines. These technological innovations in the field of Internet impacts millions of people worldwide.

Email and its by-products have become a common and cheaper mode to reach millions of end users for business houses and others. At the same time criminals use spam to mount DDoS, Man in the middle attacks to cripple network servers and other resources which are handling online business. Phishing is a kind of social engineering technique based attack luring thousands of end users to lose their money and peace. The phishing attack targets online business such as banking, financial sectors and others.

Comment spam refers unsolicited messages or trackbacks or ping pongs posted in a web page, blog, forum, mail archives. These are often irrelevant to the main body of post. The spammers are inserting their target URL to attract more traffic to their website or blog. Spammers use comment spam as an agent to list their intended website in search engines order list. There are lot of semi, full automated tools to handle these spam, but completely eradicating the comment spam attack is cumbersome. Even though email and comment spam are similar in many ways, there are lot of differences between attacks & defense mechanisms. Comment spam reaches limited number of audiences based on the traffic in the particular website, blog, and forum. The traffic in a particular website or blog post is based on its popularity.

At a time an attacker can send millions of spam mails using automated tools such as Phasma Email Spoofer, Bulk Mailer, Aneima, Avalanche, Euthanasia etc.,. These tools are highly sophisticated and are capable to bypass the defense methods of anti spam software. Even the email service providers are using lot of technologies to stop spamming, the spammers are always doing better job than anti spammers. Since the comment spam posting is not private like email spam, the content of comment spam reaches end users through search engines. The success of comment spam is visible since the attacker can see the results immediately. If the comment spam attack is not successful, the attacker can change the method of attack. But this is not possible in email spamming. There are many tools freely available in the market to send email spam. These tools can send spam without manual intervention to many users. But most comment spammers need more manual effort to post comment spam. Comment spamming tools are not sophisticated like email spamming tools. The mail administrators have lot of tools to identify the spam such as content filters, DNSBL, SURBL, rDNS etc. These kind of facilities are not available to defend comment spamming.

The rest of the paper is organized as follows. Section 2 is dedicated to related work, mostly previously proposed methods for blog spam tools based on Bayesian algorithm. Algorithm of the proposed mechanism & working principle of the tool explained in section 3. In Section 4, the experimental topology and results are explained. The conclusion of this paper is presented in the final section.

# 2. RELATED WORK

[1] Authors extensively studied the characteristics of spam and the technology used by spammers. In order to counter anti-spam technology, spammers change their mode of operation.

These evaluations help us to enhance the existing anti spam technology and thereby help us to combat spam effectively. They observed that spammers use software tools to send spam with attachment. The main features of this software are hiding senders' identity, randomly selecting text messages, identifying open relay machines, mass mailing capability and defining the duration of spamming. Spammers do not use spam software to send spam without attachment. Another important revelation of their study is that, relatively age old heavy user's email accounts attract more spam than old light user's mail accounts. Relatively new email accounts are not attracting spam except attacks caused by virus, worms & malware. We believe that this analysis could be useful to develop more efficient anti spam techniques for spam including blog comment spam.

[2] The authors presented an approach to classify the blog link comment spam by comparing the language models used in blogs and its comments post. According to their approach, they identified three language models such as language model used in original blog post and two language models used in comment post. The comment posts two language models are the language models used in comment area and the language model used in the page linked by the comment. To compare these language models, the authors used Interpolated aggregate smoothing and probabilities which are calculated by using maximum likelihood models and Jelinek mercer smoothing. Here this approach is applied for both contents of the comment and linking blog contents. The authors exploited the weakness of the spammers trying to create link between two sites that have no semantic relation. Since the spammers enhance their methods quickly, this approach will become less effective. [3] Blog spam- A Review. The author presented an in-depth analysis of spam in blog in particular as seen at major blog service TypePad. Author used two open source statistical email anti-spam solutions to classify blog comment spam and experiment results shows that they are effective. Their study reveals that comment spam traffic volume has gone down during the weekends. [4]Authors proposed a filter based on text of the junk mail and domain specific features of the problem. In order to design filters, the authors used Bayesian Network theorem applied to a classification task. By using Bayes theorem, they were able to make use of probabilistic learning methods in conjunction with a notion of differential misclassification cost. According to the authors, their methodology is well suitable for spam filters. Even though their study is very old and not suitable for current trends, they proved that the Bayesian approach is useful to design spam filters. In olden days, the domain of the email sender provides additional facility to identify the characteristics of the email such as there is no spam from .edu domain. But this concept is not acceptable these days. The attacker can use any domain by brute force method to spam end users. The authors used approximately 35 phrases to identify spam emails such as "FREE!", "only $" etc.,

## 3. OPERATIONAL METHODOLOGY OF COMMENT SPAM

In most cases, to post a comment the users are asked to write their name and email addresses. The blog itself will identify the location of the comment writer by recording the user's ip address. The comment posted by the end users will be displayed immediately in dynamic blogs, forums and discussion boards.  In these dynamic blogs, forums & web pages, the spammers take the freedom to attack the service by spamming. In contrast some blogs, forums and news archives ask users to get a login and password by submitting identity proof or some private details. Such kind of registration process increases the burden of end users. These kind of static blogs do not attract more traffic to their comment area due to reluctant of end users to divulge their identity and other details. Topix is one of the worlds largest community news site supporting dynamic comment postings from the users. Topix gives freedom to end users to kill the problematic postings. Approximately topix is handling 120000 comments per day [monitored from $20^{th}$ to $30^{th}$ September 2008]. The advantage of the topix is that, the postings will be displayed immediately without joining the community. Blogherald is similar to topix in handling dynamic postings and is supported by wordpress plugin Akismet. Livebooking Network is another dynamic comment post supporting service provider, powered with deletion of comment duplication. WordPress is an anti comment spam service provider, which also works with other blogging systems, giving more bloggers much needed spam protection. Wordpress plugin Akismet has received more than 7 billion blog spam and says that 88% of all comments spam. Akismet is handling 1902 000 average spam per day approximately [monitored on $20^{th}$ to $30^{th}$ September 2008]. Live Journal is an online journal with emphasis on user interaction. My space is similar to Live journal and both are free service providers. Weblog & Vox allow comments from registered users and comments are moderated and will not appear until the administrator has approved them. Google blogger needs users name, password and word verification test to post comments. Typepad also requires login and password to post comments in the comment areas. Twitter,Jaiku,Qik & Facebook are other major spam vulnerable blog service providers. There are more than hundred such service providers and millions of end users are publishing their stories on blogs

### 3.1 Mechanism of Proposed Comment Spam Tool

To prevent spams from sites which allows commenting, blog spam filter can be used. This supports a reliable protection to the blog,forum, mail archive & guestbooks owners. It works with detecting spam words in a comment and finding the probability whether the comment is spam or not. To find the probability of any spam the bayesian algorithm is used. Incoming comment and earlier comments stored in database are used to derive this formula. To by pass spam filters, spammers use punctuations, capital letters in between small letters and special characters to modify the meaning to the filters. To avoid this, the words that are sent from users are counted and unneccesary words like punctuation and special characters elimininated. After this elimination process, the entire user comment should be stored in database in lower case format.

Bayesian spam filtering has become a popular mechanism to distinguish spam from legitimate email. We are using Bayesian algorithm to identify the comment spam & designed a software tool to defend comment spam attacks. Bayes' theorem, in the context of spam, says that the probability that an email is spam, given that it has certain words in it, is equal to the probability of finding those certain words in spam email, times the probability that any email is spam, divided by the probability of finding those words in any email:

Pr(spam|words)=Pr(words|spam)Pr(spam)/Pr(words)……………………..(1)

Here we are going to use (1) it for comment spam. Since we are going to use this formula for comment, we have changed according to our requirement. The probability that a comment is spam, given that it has certain words in it, is equal to the probability of finding those certain words in comment spam, times the probability that any comment is spam, divided by the probability of finding those words in any comment :

Pr(comment spam|words)=Pr(words| comment spam)Pr(comment spam)/Pr(words)

## ALGORITHM OF THE SOFTWARE TOOL

The following algorithm shows the clear structure of our blog comment spam blocker methodology**.**

Step 1: The system asks user to enter their name and email addresses. The entered names and email addresses are verified for its validity before connecting to the database and web page.

Step 2: The users IP address shoud be noted and stored in the database.

Step 3: Unwanted characters like punctuations and special characters should be removed.

Step 4: All words should be converted to lowercase letters.

Step 5: Unnecessary words should be removed & total number of words counted.

Step 6: Identify the probability of incomming comment based on Bayesian algorithm with the help of data in database & user input.

Step 7: The probablity of the comment spam will be compared with threshold value. If the probablity value is equal or greater than the threshold value, the incomming comment will be rejected and stored in database for future use. Other wise the comment will be posted in the web immediately.

Step 8:If the user input is spam, the IP address of the sender will be stored in database.

Step 9:Go to step 01

Step 10: Again if there is any comment spam from the same IP address for a particular number of times,

the IP address will be blocked for a particular period
of time.

Step 11:To avoid flooding of data, the user should
be barred to send comment without interval.

## 3.2 Working Methodology

We selected Bayesian algorithm to derive a formula to detect whether the given comment is
spam or not. We derived this formula carefully to avoid false positive and false negative
comments. A legitimate comment may contain a couple of bad words but this does not mean that
the comment is a spam. Also, a spam comment may contain some good words and this does not
mean that the comment is legitimate. Therefore, a probability mechanism is designed to
distinguish incoming comments. We derived this formula based on Bayesian algorithm to block
incoming comments from the end-users. To implement this algorithm, the administrator should
store sample spam words in the database. The incoming comment will be used along with the
existing data in the database.

The first 5 steps mentioned in this algorithm are basic programming techniques, so we are not
going to explain them here. We are going to deal directly with the mechanism to identify the
spam comment. Our first step is to establish a connection between the database and our comment
page. İn step 1, The system asks users to enter their name & email address. The entered names
and email addresses are verified for its validity before connecting to the database and web page.
Without proper name & mail address, users are not allowed to enter a comment. To support
dynamic user participation,the users are encouraged to post their comment without registering
into the local host. After submitting, the page is directly connected to the database.

In Step 2,the users IP address shoud be noted and stored in the database. İn the next step,
unwanted characters like punctuations and special characters should be removed.
Spammers mix lower case, uppercase characters and punctuations to bypass the filters. To
prevent this, all incoming user's comments will be free from punctuations and all words will be
converted to lower case. In step 4, comments should be eliminated from unnecessary words.
Those words are not counted as spam or non-spam. However, they are counted from the spam
filter. To increase the performance of the spam filter, unnecessary words should be ignored.
These words must be defined by the programmer. Selection of those words depends on the
programmer. In Step 5 of the algorithm, unnecessary words should be removed & total number of
words counted. The unnecessary words include pronouns
like'am','is','are','he','she','it','you','we','they','i','have','has','had','and','us','do','does','did','was','were','
a','an','in','on','the','to','but','of','from','them','also','their' will be removed. The administrator can
define any such words by himself. This part of the program shows the indication of
unnecessary words and elimination of those words from the comment text.

```
$gereksizler=array('am','is','are','he','she','it','you','we','they','i','have','has',
'had','and','us','do','does','did','was','were','a','an','in','on','the','to','but','of','
from','them','also','their');

$dizi=explode(" ",$comment_db);

foreach($dizi as $kelime){

  $kontrol=0;

  foreach($gereksizler as $gereksiz){

    if($gereksiz==$kelime)

      $kontrol++;

  }

  if($kontrol==0)

    $kelimeler[$kelime]++;

}

$sum55=0;

foreach($kelimeler as $key=>$value){

fwrite($file2,$key." = ".$value." ");

$sum55=$sum55+$value;

$sor="insert into kelimetablosu2(  form_id, kelime, adet) values ( '1',
'$key','$value' )";

$register= mysql_query( $sor );}
```

Fig. 1 Algorithm: Elimination of unnecessary words from the comment

Step 6: Identify the probability of incomming comment based on Bayesian algorithm with the help of data in database & user input. In step 6, defining of the spam words is the most important part. Like unnecessary words, spam words should be defined by the programmers. Programmers can add spam words to the database, so that the program can find out the spam words to block easily. The following section shows example of spam word definition. The spam word replication is also an important factor to calculate the probability. The number of repeated spam words should be included as a new spam word in the calculation.

This part denotes the defining the spam words and theirs repetition times. This repetition numbers are quite important, because it directly effects to the probability of an email is spam value.

```
$bad_words=array( 'idiot' , 'stupid' , 'bad' , 'awful' , 'terrible' ,
'disgusting' , 'silly' , 'freak' , 'fool' ,'rubbish' ,'ugly' );

$count=str_word_count($comment_db);

$dizi=explode(" ",$comment_db);

    foreach($dizi as $kelime){

    foreach($bad_words as $bad){

   if($bad==$kelime)

     $bad_count++;}

}

fwrite($file2," bad words in our mail:");

fwrite($file2,$bad_count);
```

Fig. 2 Algorithm: Defining spam words and theirs repetition times

To identify the probability of a comment being spam, the stored spam word database is used as a reference factor in Bayesian algorithm.

Pr (spam| words) = Pr (words | spam) * Pr (spam) / Pr (words)......................(2)

The modified structure of Bayesian Algorithm is shown below.

Pr (comment spam| words) = Pr (words | comment spam) * Pr (comment spam) / Pr (words)…………………………………………………………………...... (3)

In detail, the Pr (words | Comment spam) can be calculated as follows

Pr (words | comment spam) =Number of spam words in user input/ Total Number of words in user input………………………………….…….....………… (4)

Pr (comment spam) can be calculated as follows from the database

Pr (comment spam) = total number of Comment spam / total number of comments…
.....................................................................................................(5)

Pr (words) can be calculated as follows

Pr (words)=total number of spam words detected in legitimate comment as well as spam comment / total number of comments…………………………………(6)

These formulas can be explained in words as, Pr(words/spam) is equal to spam words in the current comment divided by the total number of words in this comment. Total number of words doesn't include unnecessary words mentioned before.

Pr (spam) is equal to finding total comment spam stored in database divided by the total number of comments entered by the users. This value depends on past entries stored in the database of the web host. Pr (words) is equal to total number of spam words detected in any comment divided by the total number of comments. This includes the spam word detected in a legitimate comment as well as a comment spam. The numerator shows that the number of spam words is encountered in any comment.

If the probability of being a spam comment is greater than the threshold value, the comment will be classified as spam or else it will be posted online. The algorithm is shown below.

If probability of being spam >= user defined threshold value

It will reject the incoming comment as spam

Else

It will be posted online

End if

words_spam = $bad_count/$sum55;

spam=$count2/$count1;

words=$count_true2/$count1;

spam_words=($pr_words_spam*$pr_spam)/($pr_words);

If($pr_spam_words >= 0.35){

echo "Your entity is wrong,try again";}

Fig. 3:Algorithm: Defining comment spam

To block the spam mails a threshold value should be defined. This value totally depends on the programmers and their desires. The lower threshold value will protect from spam comments. This part of the program is the core of the comment spam filter tool. As it appears from the program, any probability value which is greater or equal to the threshold value is blocked. The user is warned that the comment is spam & it will be stored in the database, but not displayed. The admin can check the database and make a decision to display or not, if necessary. Also, the probability of the comment text can be lower than our threshold value. In this time, the comment will be displayed immediately. If the probability of being a spam comment is very high compared to threshold value, the entire comment statement can be added to the spam word database.

## 4.EXPERIMENTAL TOPOLOGY & RESULTS

To do the experiments, we setup a web page with database connectivity. We used WAMP paradigm to setup this webapge with database connectivity WAMP stands for Windows, Apache, MySQL, PHP stack that can be used as a platform for web applications [5]. We used MySQL for backend and PHP as a frontend and Apache2triad for inputting users data to the SQL database. Before starting to filter the spams, we need a database and a web page that are connected to eachother. For database part MySQL is selected and PHP is used for coding the Spam filter. For graphical MySQL database interface, we used PHPMyAdmin. Apache2triad is selected to send users information with their comments and theirs relation to the database. We deployed our web software in windows server using Apache2Triad to execute PHP codes. WAMP is a package of independently-created programs installed on computers based on PC operating systems. The interaction of these programs enables dynamic web pages to be served over a computer network, such as the internet or intranet.Another part of this project is to identify proper threshold value to provide a good solution to the problem. Before the threshold value is obtained, some questions should be answered. For what purpose should this spam filter be used? What kind of security level is needed? After those questions are answered, the threshold value can be obtained for reliable service.

## 4.1 Experimental Results

We tested our spam filtering tool with the help of students. We asked a group of our students to send comment spam to the particular web page loaded with our comment spam blocking tool. We changed the threshold value from 0.1 to 0.99 and monitored the output. For all experiments, we asked our 50 students to enter comments in a particular page. For the first session we set up the threshold value=0.1. If the probability of the comment being a spam is equal or greater than .1, the incoming comment will be classified as spam and it will be blocked from being displayed on the web page. In this case, our software identified all comment spam (100%) entered by the users but it also classified legitimate mail as spam. The approximate number of false positives is 20%. Here some of the comments got one or two spam words, which also got blocked as a spam comment. At the same time the number of false negative is 0%. We conducted the same kind of experiments with the threshold value up to 0.99 as shown in the table. From the experiments we identified that if we use a low threshold value, the tool blocked all spam comments but the rate of false positive is high. If we use a high threshold value the false negative is increasing and the spam identification is also not working well. The performance of the software is similar to major spam killer tools available. But for the mid range threshold value from 3.5 to 5.5, the comment spam blocking tools performs well. In that range the tool identified all spam comments. The number of false positive or false negative is almost 0 or negligible.The test results are given below.

## Table 1.Experimental results

| Threshold value | Spam identified (%) | False positive | False Negative |
|---|---|---|---|
| .1 | 100 | 20% | 0 |
| .2 | 100 | 19 | 0 |
| .3 | 100 | 9 | 0 |
| .4 | 100 | 7 | 0 |
| .5 | 100 | 1 | 0 |
| .6 | 91 | 0 | 9 |
| .7 | 85 | 0 | 15 |
| .8 | 80 | 1 | 19 |
| .9 | 71 | 0 | 19 |

Selecting threshold value is a key asset to identify the spam comment. The administrator can block all incoming comment spam by fine tuning of the spam filtering tool. An empirical study of spam & spammers characteristics plays an important role in selecting & adding spam word for the database table.

The length of the comment in our blog or web page is not limited like most of the major blog, forum and message archive services in the Internet. The size of the spam word database is not limited to give space for more accurate performance of the tool.

The size of our software tool is much less, so this tool can be loaded in any level of server or web page handling comments. Since the line of code is very less, it will not increase the access time of the web page.

The sample output is shown below. The web page accepts comment, name, surname, age and e-mail areas. After entering the fields, the user should click the submit button. Now, we will see the results after submitting. As you can see in the Fig.2. the comment has been displayed. We fixed the threshold value equal to " 0.35 ". because, the probability of a comment being a spam is " 0.320819112628 ". This comment contains 3 spam words which is defined by the administrator (bad,disgusting,stupid). The spam word "bad" repeats two times, so, the probability of this comment is calculated by using 4/12 with other data. The number of words after eliminating unnecessary words are 12.

COMMENT --> yesterday I went to a restaurant it was a bad experience food were disgusting and also waiters were stupid but the view was not bad NAME--> Test SURNAME -->Test AGE--> 15E-MAIL-->test@test.com Total entries:358count_true: 295 Total words:3335 Total bad mails:286 pr_spam_words: 0.323163841808 Your comment has been sent

**Fig.3** Output after submitting

Therefore, any comment which has the lower probability value than the threshold value can easily pass the filter and will be posted in the web page immediatly.

In the second example,the comment contains 7 spam words. (Awful, fool, bad, rubbish, terrible, disgusting). Its probability as being a spam is "0.484797297297". Since this value is greater than the threshold value, the user comment is blocked by the tool.

## 5. CONCLUSION

Spam is a simple methodology to spread commercials through email, blogs, forums, email archives and instant messengers. Virus, worms, and trojon deployment through spam is forcing business houses to spend millions of dollars each year to protect their resources from these attackers. Comment spam is a type of spam annoying end users and wasting bandwidth. There were many methods proposed in the past to stop but nothing worked properly against spammer's technology. In this paper we proposed an approach based on Bayesian algorithm. The total line of code for this software is much less. It can be deployed in any level of server or web page. The experimental results show that our software tool works well to protect users from comment spammers. Our test results show that the fine tuning of the threshold value & continuous updating of the spam words can play an important role to completely eliminate the comment spam. The main problem with this approach is that it requires constant updating since spammers constantly find new ways to by pass any filter.